# Social Force in Pedestrian Crowd

**Peng Wang** **Xiaoda Wang**

***Abstract—***

This paper provides a new perspective to understand existing controversy on the social force model. These issues include that the social force disobeys Newton 3rd Law, oscillation phenomenon when one agent is approaching another as well as some questions on the faster-is-slower effect. From the perspective of physics these problems seems difficult to explain. This paper provides a new perspective to understand these issues. We introduce a new concept of desired interpersonal distance to explain how the social force is generated from conscious mind of human. Although the social force disobeys Newton 3rd Law, the whole model is exactly within the Newton Laws to characterize pedestrian motion. The oscillation phenomenon may exist in non-physics entity (i.e., desired velocity and desired interpersonal distance) rather than physics entity (i.e., actual velocity and actual distance), and such oscillation is mitigated by treating non-physics entity as variable rather than constant. Very interestingly, the desired velocity represents the motivation level of pedestrian motion, and the faster-is-slower effect is thus explained by Yerkes–Dodson law, explaining how motivation level could improve or impair human performances in a collective sense. This inverted-U effect is further studied with a falling-down model and the numerical testing is exhibited by using FDS+Evac.

## I. About the Social-Force Model

Modeling of pedestrian motion has interested scientists in various disciplines. As computer technology has been developed over recent years, a major approach to study pedestrian behavior is describing individual motion by a mathematical model and using computer programs to simulate how collective motion emerges from individual interactions. A well-known model in this area is called social-force model, and it was introduced by physical scientists (Helbing and Molnar, 1995; Helbing, Farkas and Vicsek, 2000; Helbing et al., 2002). The model characterizes individual pedestrian motion based on Newton dynamics. Very interestingly, certain psychological factors are abstracted in model parameters such as “desired velocity” and “social force,” and these concepts are not physical entities because they describe people's opinions in human mind. Thus, the model is not typically within the scope of physics study, but in an interdisciplinary field, and an important question is whether such innovation is consistent with physics laws and psychological principles on human.

As the social-force model was initially introduced, it was applied to either many-particle simulation in statistics physics or pedestrian simulation in traffic systems (Helbing, 2001). Thus the individual model is either for “particle” or “human” in a sense. However, in the past two decades a great number of computer engineers have applied this model to pedestrian simulation, and the model is now widely accepted for pedestrian modeling, not for particles. This situation bring some critical questions to the model. For example, the anisotropic social-force disobeys the Newton Third law and this is questionable because human pedestrian motion are definitely within Newton Laws. All the forces implemented for human motion should

Peng Wang previously studied in the Department of Electrical and Computer Engineering, University of Connecticut, Storrs, USA. Parts of the research work were conducted in his graduate program in University of Connecticut.

follow Newton 3rd Laws. Thus, in this contribution we will try to provide a new perspective to understand the model, and the model will be renewed mainly referring to the social force. Very importantly, a new concept of desired interpersonal distance is essentially introduced, by which we will explain the model agrees with Newton 3rd Law for pedestrian motion. Based on the renewed model we further discuss the oscillation phenomenon and explain the faster-is-slower effect by Yerkes-Dodson law. Before we go to such detailed discussion, let us briefly review the mathematical description about the social force model as below.

The social-force model presents psychological forces that drive pedestrians to move as well as keep a proper distance with others. In this model an individual's motion is motivated by a self-driving force $\boldsymbol{f}_i^{self}$ and resistances come from surrounding individuals and facilities (e.g., walls). Especially, the model describes the social-psychological tendency of two individuals to keep proper interpersonal distance (as called the social-force) in collective motion, and if people have physical contact with each other, physical forces are also taken into account. Let $\boldsymbol{f}_{ij}$ denote the interaction from individual $j$ to individual $i$, and $\boldsymbol{f}_{iw}$ denote the force from walls or other facilities to individual $i$. The change of the instantaneous velocity $\boldsymbol{v}_i(t)$ of individual $i$ is given by the Newton Second Law:

$$m_i \frac{d\,\boldsymbol{v}_i(t)}{dt} = \boldsymbol{f}_i^{self} + \sum_{j(\neq i)} \boldsymbol{f}_{ij} + \sum_{w} \boldsymbol{f}_{iw}$$

$$m_i \frac{d\,\boldsymbol{v}_i(t)}{dt} = \boldsymbol{f}_i^{self} + \sum_{j(\neq i)} (\boldsymbol{f}_{ij}^{soc} + \boldsymbol{f}_{ij}^{phy}) + \sum_{w} (\boldsymbol{f}_{iw}^{soc} + \boldsymbol{f}_{iw}^{phy}) \tag{1}$$

where $m_i$ is the mass of individual $i$. Furthermore, the self-driving force $\boldsymbol{f}_i^{self}$ is specified by

$$\boldsymbol{f}_i^{self} = m_i \frac{\boldsymbol{v}_i^0(t) - \boldsymbol{v}_i(t)}{\tau_i}\ , \tag{2}$$

This force describes an individual tries to move with a desired velocity $\boldsymbol{v}_i^0(t)$ and expects to adapt the actual velocity $\boldsymbol{v}_i(t)$ to the desired velocity $\boldsymbol{v}_i^0(t)$ with a characteristic time $\tau_i$. In particular, the desired velocity $\boldsymbol{v}_i^0(t)$ is the target velocity existing in one's mind while the actual velocity $\boldsymbol{v}_i(t)$ characterizes the physical speed and direction being achieved in the reality. The difference of $\boldsymbol{v}_i^0(t)$ and $\boldsymbol{v}_i(t)$ implies the gap between the human subjective wish and realistic situation, and it is scaled by a time parameter $\tau_i$ to generate the self-driving force. This force motivates one to either accelerate or decelerate, making the realistic velocity $\boldsymbol{v}_i(t)$ approaching the desired velocity $\boldsymbol{v}_i^0(t)$. This mathematical description of the self-driving force could be dated back to the Payne-Whitham traffic flow model (Payne, 1971; Whitham, 1974), when the stop-and-go wave is studied for vehicle traffic problem. This assumption is also used in several other pedestrian models such as Centrifugal Force Model (Chraibi, 2011). Sometimes $\boldsymbol{v}_i^0(t)$ is rewritten as $\boldsymbol{v}_i^0(t) = v_i^0(t)\boldsymbol{e}_i^0(t)$, where $v_i^0(t)$ is the desired moving speed and $\boldsymbol{e}_i^0(t)$ is the desired moving direction. In a similar manner, we also have $\boldsymbol{v}_i(t) = v_i(t)\boldsymbol{e}_i(t)$ where $v_i(t)$ and $\boldsymbol{e}_i(t)$ represent the physical moving speed and direction, respectively.

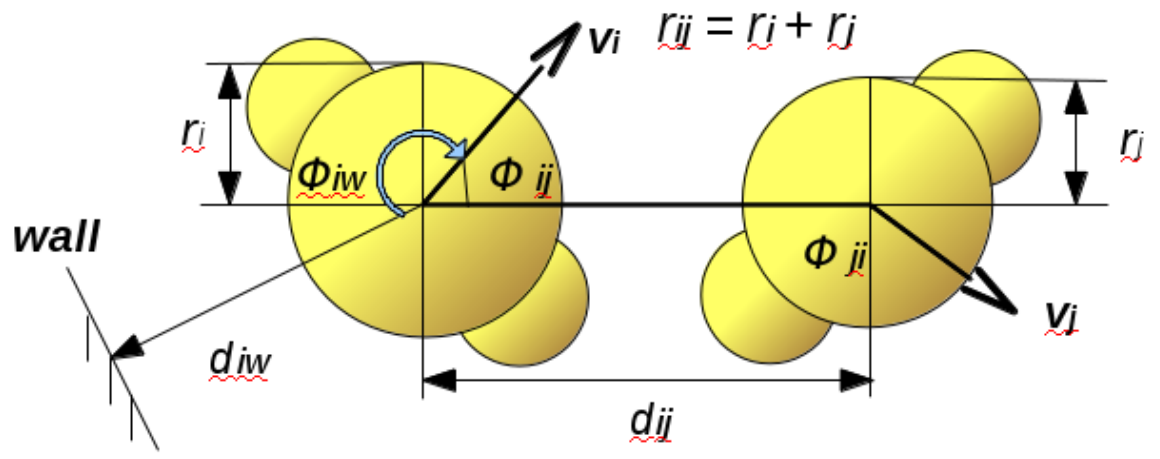

**Figure 1. A Schematic View of Two Pedestrians (Equation 3 and 4): The distance of pedestrians *i* and *j* is denoted by $d_{ij}$ and the sum of their radii is given by $r_{ij}$. The distance to the wall is denoted by $d_{iw}$.**

The interaction force of pedestrians consists of the social-force $\boldsymbol{f}_{ij}^{soc}$ and physical interaction $\boldsymbol{f}_{ij}^{phy}$. i.e., $\boldsymbol{f}_{ij} = \boldsymbol{f}_{ij}^{soc} + \boldsymbol{f}_{ij}^{phy}$. The social-force $\boldsymbol{f}_{ij}^{soc}$ characterizes the social-psychological tendency of two pedestrians to stay away from each other, and it is given by

$$\boldsymbol{f}_{ij}^{soc} = A_i \exp\left[\frac{(r_{ij} - d_{ij})}{B_i}\right] \boldsymbol{n}_{ij} \quad \text{or} \quad \boldsymbol{f}_{ij}^{soc} = \left(\lambda_i + (1-\lambda_i)\frac{1+\cos\phi_{ij}}{2}\right) A_i \exp\left[\frac{(r_{ij} - d_{ij})}{B_i}\right] \boldsymbol{n}_{ij} \tag{3}$$

where $A_i$ and $B_i$ are positive constants, which affect the strength and effective range about how two pedestrians are repulsive to each other. The distance of pedestrians $i$ and $j$ is denoted by $d_{ij}$ and the sum of their radii is given by $r_{ij}$ . $\boldsymbol{n}_{ij}$ is the normalized vector which points from pedestrian $j$ to $i$. The geometric features of two pedestrians are illustrated in Figure 1. In practical simulation, an anisotropic formula of the social-force is widely applied where Equation (3) is scaled by a function of $\lambda_i$. The angle $\varphi_{ij}$ is the angle between the direction of the motion of pedestrian $i$ and the direction to pedestrian $j$, which is exerting the repulsive force on pedestrian $i$. If $\lambda_i = 1$, the social force is symmetric and $0 < \lambda_i < 1$ implies that the force is larger in front of a pedestrian than behind. This anisotropic formula assumes that pedestrians move forward, not backward, and thus we can differ the front side from the backside of pedestrians based on their movement. Although the anisotropic formula is widely used in pedestrian modeling, it also brings a controversial issue that the anisotropic formula of social force disobeys Newton's 3rd law.

The physical interaction $\boldsymbol{f}_{ij}^{phy}$ describes the physical interaction when pedestrians have body contact, and it is composed by an elastic force that counteracts body compression and a sliding friction force that impedes relative tangential motion of two pedestrians. Both of them are valid only when $r_{ij}>d_{ij}$. In Helbing, Farkas and Vicsek, 2000 the interaction force is repulsive. The model may also include an attraction force in its original version (Helbing and Molnar, 1995, Korhonen, 2015). The interaction of a pedestrian with obstacles like walls is denoted by $\boldsymbol{f}_{iw}$ and is treated analogously, i.e., $\boldsymbol{f}_{iw} = \boldsymbol{f}_{iw}^{soc} + \boldsymbol{f}_{iw}^{phy}$. Here $\boldsymbol{f}_{iw}^{soc}$ is also an exponential term and $\boldsymbol{f}_{iw}^{phy}$ is the physical interaction when pedestrians touch the wall physically. In particular, the exponential term is given by

$$\boldsymbol{f}_{iw}^{soc} = A_{iw} \exp\left[\frac{(r_i - d_{iw})}{B_{iw}}\right] \boldsymbol{n}_{iw} \quad \text{or} \quad \boldsymbol{f}_{iw}^{soc} = \left(\lambda_{iw} + (1-\lambda_{iw})\frac{1+\cos\phi_{iw}}{2}\right) A_{iw} \exp\left[\frac{(r_i - d_{iw})}{B_{iw}}\right] \boldsymbol{n}_{iw} \tag{4}$$

where $A_{iw}$ and $B_{iw}$ are positive constants. The distance of pedestrians $i$ and the wall is denoted by $d_{iw}$ and $\boldsymbol{n}_{ij}$ is the normalized vector which points from the wall to pedestrian $i$. The angle $\varphi_{iw}$ is the angle between the direction of the motion of pedestrian $i$ and the direction to the wall, which is exerting the repulsive force on pedestrian $i$.

By simulating many such individuals in collective motion, several scenarios in crowd movement were demonstrated, and one is called the "faster-is-slower" effect. This scenario was observed when a crowd pass a bottleneck doorway, and it shows that increase of desired speed (i.e., $|\boldsymbol{v}_i^0(t)|$) can inversely decrease the collective speed of crowd passing through the doorway. Another paradoxical phenomenon is called "freezing-by-heating," and it studies two groups of people moving oppositely in a narrow passageway, and the simulation shows that increasing the fluctuation force in Equation (1) can also cause blocking in the counter-flow of pedestrian crowd. Other spatio-temporal patterns include herding effect, oscillation of passing directions, lane formation, dynamics at intersections and so forth.

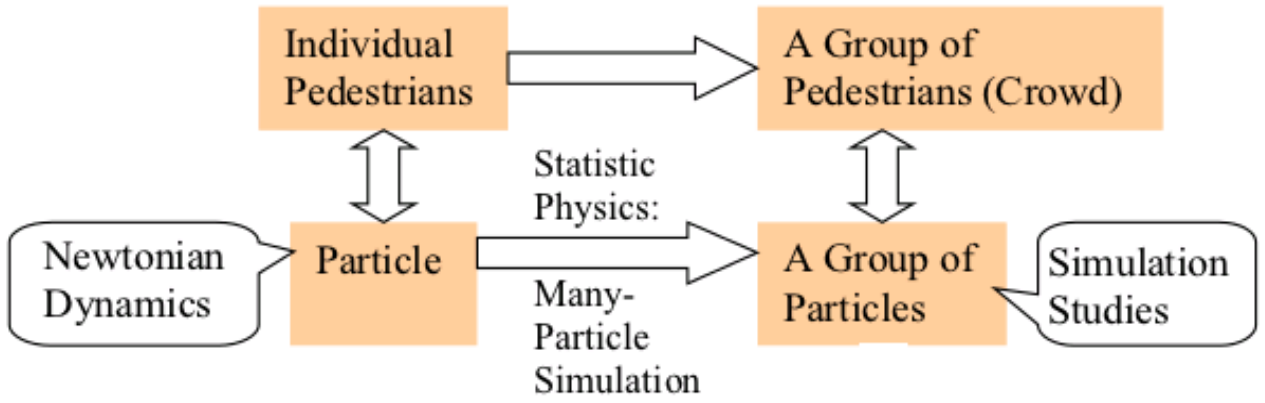

**Figure 2. From Particle Dynamics to Crowd Simulation:**
**The Framework of Many-particle Simulation in Helbing, Farkas, and Vicsek, 2000**

In the past decade, the social-force model has generated considerable research on evacuation modeling (Helbing and Johansson, 2010), and it has been integrated into several egress simulators, such as Fire Dynamics Simulator with Evacuation (Korhonen et. al., 2008; Korhonen and Hostikka, 2010; Korhonen, 2019), Pedsim, SimWalk, MassMotion (Oasys, 2018), VisWalk, Maces (Pelechano and Badler, 2006) and Menge. The model has been partly validated based on data sets from real-world experiments. The method of validation involves comparing the simulation of the model with associated observations drawn from video-based analysis (Johansson, Helbing and Shukla, 2007; Johansson et al., 2008).

A criticism about the model, as mentioned before, is that the anisotropic formula of social force disobeys Newton's 3rd law. If the model does not obey Newton 3rd law, it becomes questionable in field of physics studies. Another problem about the social force is the dilemma of choosing proper parameters to avoid both overlapping and oscillations of the moving pedestrians (Chraibi, et al., 2011). Certain scenarios about oscillation of walking behavior are not realistic in the physical world. From the perspective of physics these problems are difficult to deal with. However, as mentioned previously, human motion is self-driven and self-adapted, and it is subject to both physical laws and psychological principles. Psychological study will give a new perspective to understand the model and provide us with a new angle to understand the problems. This is why we want to bridge the gap between psychological studies and physical explanation about this model.

## II. Social-Force and Newton Third Law

This section provides a psychological perspective to reinterpret the social-force, and the concepts of interpersonal distance, proxemics and stress are critically involved. Very importantly we introduce a new concept of desired distance in the social force, which is the counterpart of the desired velocity in the self-driving force. With this new concepts we further discuss whether the model is consistent with Newton 3rd Law.

### A. Interpersonal Space and Social-Force

The concept of social force may originate from a concept by Lewin (Lewin, 1951), where social fields or social forces were first introduced to guide behavioral changes of people. In soci-psychology, such social fields or forces are representations of certain social norms that regulate people's behavior when one individual interacts with others. In a spatial sense such social norms refers to how people use their interpersonal distance, and it refers to proxemics in psychological theories.

Interpersonal distance refers to a theory of how people use their personal space to interact with surrounding people. In Hall, 1963 the theory was named by proxemics, and it was defined as "the interrelated observations and theories of man's use of space as a specialized elaboration of culture." Proxemics suggests that we surround ourselves with a "bubble" of personal space, which protects us from too much arousal and helps us keep comfortable when we interact with others. Psychologically, proximal distance origins from basic human instincts that we define a personal space to get a sense of safety. People normally feel stressed when their personal space is invaded by others. There are four interpersonal distances mentioned in proxemics theory (Hall, 1963; Hall, 1966): intimate (<0.46m), personal (0.46m to 1.2m), social (1.2m to 3.7m), and public (>3.7m), and each one represents a kind of social relationship between individuals. Here we highlight two issues about proxemics as below.

(a) The interpersonal distance is object-oriented. For example, we usually keep smaller distance to a friend than to a stranger, and such distance is an indication of familiarity. As named by personal distance (0.46m to 1.2m) in proxemics, this range is widely observed as the distance to interact with our friends or family, and normal conversations take place easily at this range.

(b) The interpersonal distance reflects a kind of social norms. For example, in a crowded train or elevators, although such physical proximity is psychologically disturbing and uncomfortable, it is accepted as a social norm of modern life. Also, it is also known that the male and female commonly keep larger distance in public place in Muslim culture than other cultures. In brief although proximal distance origins from basic human instincts, it is also widely redefined in different social norms.

Proxemics implies that when the interpersonal distance is smaller than the desired, people feel stressed. Repulsion comes into being in this situation, and repulsion increases when the distance further decreases. This theory justifies the assumption of repulsive social-force in Equation (3). However, such repulsion does not depend on physical size of two people (i.e., $r_{ij}$), but the social relationship, occasions and social norms. Similar to self-driving force we suggest that it is proper to add a subjective concept of desired distance $d_{ij}^0$ in the social force, and it replaces $r_{ij}$ in Equation (3). Here $d_{ij}^0$ is the target distance that individual $i$ expects to keep with individual $j$. This distance describes the social relationship of individual $i$ and $j$ . Based on the exponential form in Equation (3), the social force is rewritten as

$$\boldsymbol{f}_{ij}^{soc} = A_i \exp\left[\frac{(d_{ij}^0 - d_{ij})}{B_i}\right]\boldsymbol{n}_{ij} \quad \text{or} \quad \boldsymbol{f}_{ij}^{soc} = \left(\lambda_i + (1-\lambda_i)\frac{1+\cos\phi_{ij}}{2}\right) A_i \exp\left[\frac{(d_{ij}^0 - d_{ij})}{B_i}\right]\boldsymbol{n}_{ij} \tag{5}$$

Similar to desired velocity $\boldsymbol{v}_i^0$, the desired distance $d_{ij}^0$ is the target distance in one's mind, specifying the distance that one expects to adapt oneself with others. The physical distance $d_{ij}$ is the distance achieved in the reality. The gap of $d_{ij}^0$ and $d_{ij}$ implies the difference between the subjective wish in one's mind and objective feature in reality. Here $A_i$ and $B_i$ are parameters as introduced before, and $\boldsymbol{n}_{ij}$ is the normalized vector which points from pedestrian $j$ to $i$. In a similar manner, an anisotropic formula of social-force is also modified in Equation (5). The social force also functions in a feedback manner to make the realistic distance $d_{ij}$ approaching towards the desired distance $d_{ij}^0$. A difference is that $\boldsymbol{v}_i^0$ and $\boldsymbol{v}_i$ are vectors while $d_{ij}^0$ and $d_{ij}$ are scalars.

A major difference between the concepts of $r_{ij}$ and $d_{ij}^0$ is that $r_{ij}$ is a physics-based concept and it is normally a constant. In contrast $d_{ij}^0$ is neither a constant nor a physics entity. In brief, $d_{ij}^0$ describes people's opinion, and it may vary as the opinion changes. For example, when crowd wait at a narrow entrance, people accept smaller interpersonal distance, and thus $d_{ij}^0$ is tuned to be a smaller value so that their repulsion is reduced. Reducing repulsion in certain conditions has been applied in crowd simulation such as FDS+Evac (Korhonen and Hostikka, 2010; Korhonen, 2019). On the other side the coding framework of social-force model is not directly affected when $r_{ij}$ is replace by $d_{ij}^0$ in computer programs When realizing the model in computer programs, $r_{ij}$ and $d_{ij}^0$ are exactly at the same position, and we can simply transform $r_{ij}$ to $d_{ij}^0$ by using $d_{ij}^0 = r_{ij} \cdot c_{ij}$ such that the existing program is extended for the new force. Here $c_{ij}>1$ is a scale factor.

Furthermore, it is feasible to modify the wall repulsion in a similar way, where $d_{iw}^0$ is the desired distance that pedestrian $i$ expects to keep with the wall, and thus the Equation (4) is modified as below.

$$\boldsymbol{f}_{iw}^{soc} = A_{iw}\exp\left[\frac{(d_{iw}^0 - d_{iw})}{B_{iw}}\right]\boldsymbol{n}_{iw} \quad \text{or} \quad \boldsymbol{f}_{iw}^{soc} = \left(\lambda_{iw} + (1-\lambda_{iw})\frac{1+\cos\phi_{iw}}{2}\right)A_{iw}\exp\left[\frac{(d_{iw}^0 - d_{iw})}{B_{iw}}\right]\boldsymbol{n}_{iw} \qquad (6)$$

In a psychological sense $d_{ij}^0$, $d_{iw}^0$ and $\boldsymbol{v}_i^0$ are both subjective concepts which exist in people's mind, and they characterize how an individual intends to interact with others and environment. As a result, the social-force given by Equation (5) and (6) and the self-driving force are both subjective forces which are generated involving one's mental activities. In a physics sense the subjective forces are generated by the foot-ground friction, which exactly obey physics laws. The social-force model thus exhibits a bridge between the physics laws and psychological principles regarding crowd motion.

## B. Social-Force and Newton 3rd Law

A common criticism about the social force model is that the anisotropic formula of social force disobeys Newton's 3rd law, and thus it becomes questionable in field of physics studies. In addition, Equation (5) implies that $d_{ij}^0$ may be different from $d_{ji}^0$. As a result, the modified social force between two individuals is also not balanced, i.e., $d_{ij}^0 \neq d_{ji}^0$ and $f_{ij}^{soc} \neq f_{ji}^{soc}$. Thus, Newton third law (actio = reactio) does not hold either for Equation (5) even if the anisotropic formula is not used.

An interesting question is why the social force itself does not obey Newton 3rd Law. The important issue is that $d_{ij}^0$ exists in human mind, and it is not a physics concept, but a subjective concept in people's mind to describe how an individual intends to interact with others. Thus, the self-driving force and social force are both entities in subjectiveness, and they are generated involving one's intention and opinion. If the anisotropic formula is used, it further takes human foresight effect into account, where each individual pedestrian is more influenced by others in front than things behind. In other words, the anisotropic formula also assumes that human has perception to the surrounding things, and such perception process is not a physics thing either, but a subjective concept involving human perception and cognition. Because the social force is not a physics concept, it does not necessarily follow Newton 3rd Law in its mathematical expression.

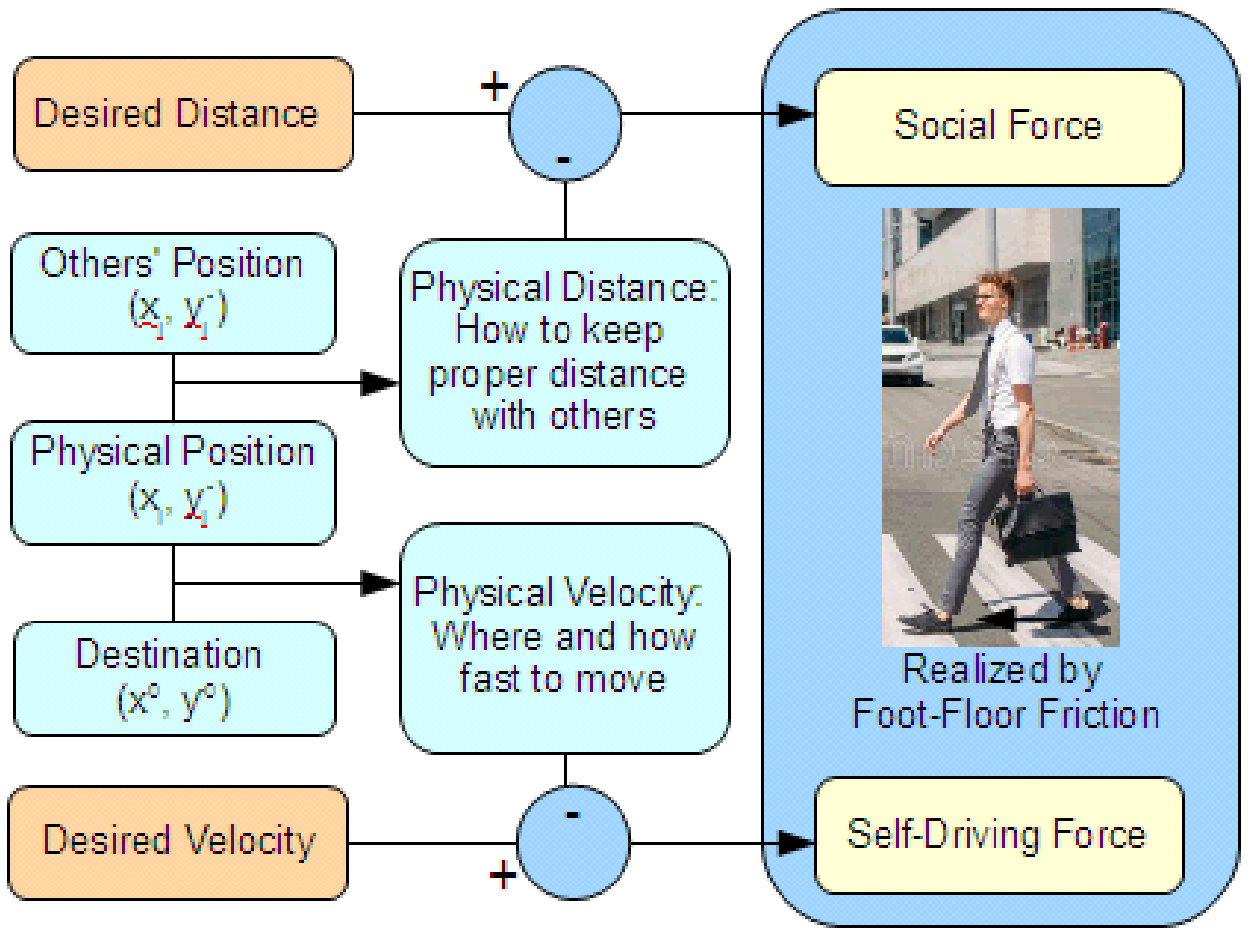

**Figure 3. Walking Behavior and Social-Force Model: When a human individual is walking, he will adjust the desired velocity and desired interpersonal distance such that he will move toward his destination while also keep a proper distance to surrounding people. Such perception and behavior is abstracted as the self-driving force and social force in the above mathematical model.**

In real-world situation two individuals never have any physical interaction if they do not touch each other physically. If a pedestrian want to keep a certain distance to anyone, he or she usually do not need to push or pull anyone, but move with their feet to adjust the distance. In other words, they use foot-floor friction to realize the "social force." Therefore, the foot-floor friction is a physics force and it is within the scope of physics study, and it is inherently consistent with Newton 3rd Law for interaction between foot and floor. The social force is not a physics concept. When the social force is realized in the physical world, it is a part of foot-floor friction.

In general any living bodies with a conscious mind has a kind of freedom of motion, and the media is an important factor to realize such motion because living bodies normally push or pull such media to realize motion as desired. For human beings the medium is the ground and we interact with the ground to move ahead. For bird or fish the media are the air or water, and they are able to fly or swim by interacting with air or water. As a result, the Newton 3rd Law holds between living bodies and such media, but not directly between two individual living bodies. In sum, the social-force model is applicable to movement of any creature like human, bird or fish, and the model is thus useful to characterize movement of any living forms with conscious mind.

Table 1. About Newton's Laws and Social Force Model

<table>
<tr><td colspan="3">$m_i \frac{d\mathbf{v}_i(t)}{dt} = \mathbf{f}_i^{self} + \sum_{j(\neq i)} (\mathbf{f}_{ij}^{soc} + \mathbf{f}_{ij}^{phy}) + \sum_{w} (\mathbf{f}_{iw}^{soc} + \mathbf{f}_{iw}^{phy})$</td></tr>
<tr><td>Self-Driving Force</td><td>$\mathbf{f}_i^{self} = m_i \frac{\mathbf{v}_i^0(t) - \mathbf{v}_i(t)}{\tau_i}$</td><td rowspan="3">These forces are generated by intentions of people and are realized by the foot-ground friction, which exactly obey Newton's Laws.</td></tr>
<tr><td>Social Force</td><td>$\mathbf{f}_{ij}^{soc} = A_i \exp\left[\frac{(d_{ij}^0 - d_{ij})}{B_i}\right] \mathbf{n}_{ij}$</td></tr>
<tr><td>Wall Repulsion (Psychological Component)</td><td>$\mathbf{f}_{iw}^{soc} = A_{iw} \exp\left[\frac{(d_{iw}^0 - d_{iw})}{B_{iw}}\right] \mathbf{n}_{iw}$</td></tr>
</table>

As shown in Table 1 foot-floor friction is the physics-based force that drive pedestrians to move, and we can consciously control this friction to decide where and how fast we move. In social force model such foot-floor friction mainly consists of self-driving force and social force as shown in Figure 3, and we assume that vectorial additivity of these force components which represents different environmental influence on the moving pedestrian. This assumption is roughly reasonable because human mind has the capability of parallel perceiving and thinking, which means people are able to adjust their interpersonal distance (i.e., social force) while still keep their destination in mind (i.e., self-driving force). If an individual is very close to another, the social force will increase such that it becomes predominant in the joint force. In contrast if an individual is not surrounded with other individuals or facilities, the self-driving force predominate the joint force such that one will just head the destination. Thus, the vectorial additivity of subjective forces is a reasonable assumption for pedestrian modeling at current stage. In particular we mainly focus on human pedestrian so that existing psychological findings can be used to interpret certain simulation results of social force model[1].

In sum the social force is a kind of interaction involving one's consciousness. It is not typically a physics thing. The social force characterizes whether an individual "desires" to be close or far away from others, and thus it inspires us to replace the individual radius by "desired interpersonal distance" to highlight this psychological effect. As a pedestrian realizes this force in the physical world, the force is part of the foot-floor friction, which exactly obey Newton's laws between the human and the ground. In other words Newton 3rd law is valid in pedestrian modeling where the social force is viewed as a part of foot-ground friction. In a word the social-force model critically exhibits a bridge between the physics laws and psychological principles regarding crowd motion, and we consider this as a major contribution of the social force model.

[1]We also note that some existing research mentioned that the social force model was also applied to movement of non-living bodies such as particles. If the social force model is used for nonliving things which interacts directly without other medium involved, one may argue that the model disobeys Newton 3rd Law. This topic is not within the scope of this paper. In this paper we only discuss how to apply social force model to the living bodies with conscious mind.

## III. Oscillating Walkers in Collective Behavior

This section will explain a controversial issue on oscillating phenomenon of pedestrian crowd, and this phenomenon is due to the dilemma of choosing proper parameters to avoid both overlapping and oscillations of the moving pedestrians (Chraibi, et al., 2011, Kretz, 2015). Certain scenarios about oscillating walkers are not realistic in the physical world. From the perspective of physics these problems are difficult to be addressed, and the problem is thus reviewed using a psychological perspective. An interesting result is that oscillation is not about the human behavior at the physics level, but the opinion in human mind.

Generally speaking, the model of pedestrian movement is commonly applied in two-dimensional space. In this section we will simplify the scenario to highlight the oscillation phenomenon, and thus limit our analysis to one-dimension space. Given $N$ pedestrians distributed uniformly in a corridor and neglect the effects of walls on pedestrians as shown in Figure 4. Here it is assumed that all the pedestrians move in a queue by giving desired distance and desired velocity, and they are modeled as simple geometric objects of constant size. The physical size of pedestrians is omitted for simplicity and desired interpersonal distance $d_0$ is highlighted.

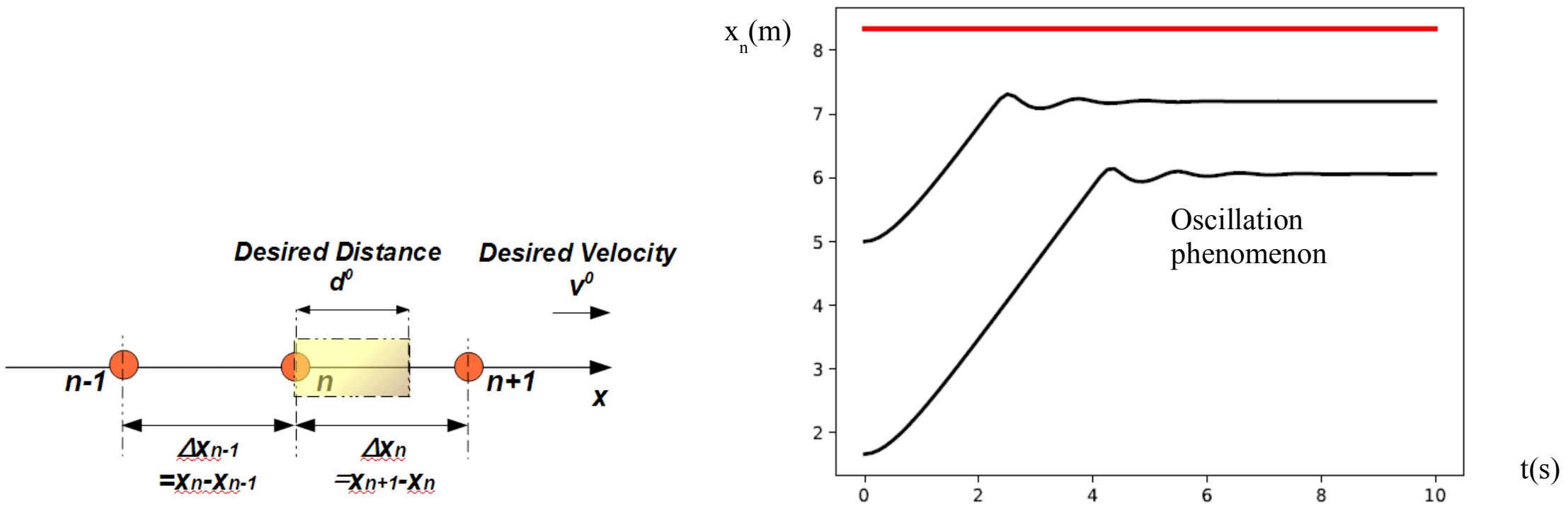

**Figure 4. Oscillating Agents in One-Dimensional Space: Three pedestrian agents are distributed uniformly as above. The physical size of agents is ignored for simplicity and the desired interpersonal distance $d^0$ is highlighted. Moreover, for interactions among agents, it is assumed that pedestrians have foresight such that pedestrian $n$ is only influenced by the pedestrian in front, not by the one behind. Oscillation is observed when agents get close sufficiently. The repulsive social force result in oscillation phenomena with $v^0$ =1.2m/s and $d^0$ =0.6m.**

Moreover, for interactions among $N$ pedestrians, we assume that pedestrians have foresight such that pedestrian $n$ is only influenced by the pedestrian right in front, not the ones behind, and this means simplifying anisotropic with $\varphi_{ij}$=0 (See Eq. 5). Another assumption is that a pedestrian is only influenced by the adjacent one in front, but not all the people, and this method significantly saves the computational effort and provides a good approximation when $d_0$ is sufficiently large. Given $\Delta x_n = x_{n+1} - x_n$, and we have the following equation by Newton Second Law.

$$\frac{d^2 x_n}{dt^2} = k_1 (v^0 - v) - k_2 \exp\left(\frac{d^0 - \Delta x_n}{k_3}\right) \tag{7}$$

where $k_1 = 1/\tau$ , $k_2 = A/m_0$ and $B = k_3$ .

In the system of $N$ pedestrians, $N$ ordinary differential equations (ODE) are formulated and it is feasible to simulate the system by a computer program. The simulation result is illustrated in Figure 4 where three pedestrians interact with $v_0$ =1.2m/s and $d_0$ =0.6m, and oscillation is observed in the testing result. A common method to study this phenomenon is to linearize Equation (7) at its equilibrium point and discuss its characteristic equation and characteristic root. In brief mathematical tools are used to tell in what conditions such oscillation occurs. In Kretz, 2015 the condition is given by $4\tau v_0 < B$ for traditional social force model. In addition there are also a number of methods to mitigate oscillation in ODE analysis. A widely-used method is adding differential element, namely the repulsion does not only depend on $d^0$-$\Delta x_n$, but also the differential of $d^0$-$\Delta x_n$ . If $d^0$ is assumed as a constant, the differential of $\Delta x_n$ is actually the relative velocity of two individual agents, and this method has been applied in several other pedestrian models (Fang et. al., 2008, Chraibi, et al., 2011).

However, the above methods only look into the problem in mathematics, but does not tell what such oscillation represents in physics. In fact several papers have clarified that oscillation of moving pedestrians is not realistic (Chraibi, 2011, Kretz, 2015). We seldom observe oscillating phenomenon when one person is approaching another in the reality. People usually have a target interpersonal distance and get there without oscillation. However, by using Equation (3), (5), (7), oscillation is observed in the numerical result, indicating that there may be something wrong with the model or assumption. Is the social force model wrong? No. We think the model is generally correct by reviewing the model in physics literature and social psychological literature, but there is a problem in applying the model to human behavior. In fact a critical problem is about the assumption that $v_0$ and $d_0$ are constants when the model is used to describe pedestrian motion. As mentioned before $v_0$ and $d_0$ are not physical entities and they actually reflect people's opinions. If $v_0$ and $d_0$ are not constant, they may change or oscillates instead of the physical velocity $v$ and distance $d=\Delta x_n$.

In another perspective the question is raised as what is the entity that oscillates. Take the distance $d_{ij}$ and desired distance $d_{ij}^0$ for example. If it is the physical distance $d_{ij}$ that oscillates, then it brings a question because we do not observe oscillating walkers in the realistic world. However, if we assume that it is possible for the desired distance $d_{ij}$ to oscillate, it means that people's opinions oscillate rather than the physical entity in the realistic world. This explanation may be reasonable because it is consistent with certain psychological theory on periodicity of stress in human behavior. More importantly, it seems that the social force model refers to a kind of mind-body problem, which are widely controversial in modern physics and philosophy. Here the interplay of mind feature (i.e, desired velocity $v_0$ and desired distance $d_0$) and body feature (i.e., physical velocity $v$ and physical distance $d$) is described within the framework the Newtonian dynamics, and they are both variables and interact with each other in a closed-loop system.

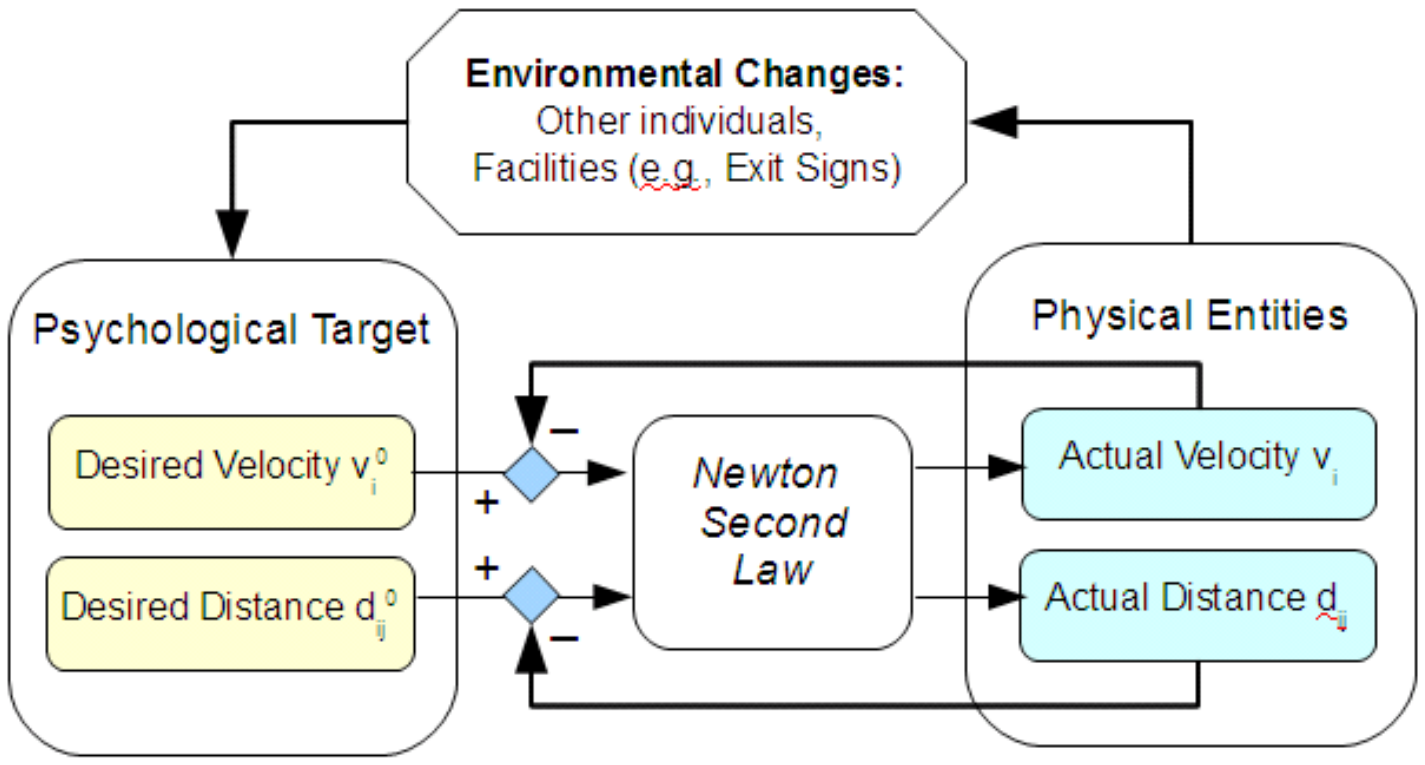

**Figure 5. Feedback Mechanism in Social Force Model: The idea of varying $d_{ij}^0$ and $v_i^0$ is also consistent with the control theory. The control targets are non-physics entities ($d_{ij}^0$ and $v_i^0$) and physics entities ($d_{ij}$ and $v_i$,) are updated to reach the targets. The interplay between the physics entities ($d_{ij}$ and $v_i$,) and non-physics entities ($d_{ij}^0$ and $v_i^0$) forms a close-loop process.**

The idea of varying $d_{ij}^0$ and $v_i^0$ is also consistent with the control theory. The control targets are non-physics entities ($d_{ij}^0$ and $v_i^0$) and physics entities ($d_{ij}$ and $v_i$,) are updated to reach the targets. The real situation is that physics entities ($d_{ij}$ and $v_i$,) are fedback to the human perception so that $d_{ij}^0$ and $v_i^0$ are also adjusted in certain conditions. The interplay between the physics entities ($d_{ij}$ and $v_i$,) and non-physics entities ($d_{ij}^0$ and $v_i^0$) forms a close-loop process as illustrated in Figure 5. However, human is not machines, and we cannot model human behavior strictly like control systems in machines. The major difference is that interplay between the physics entities ($d_{ij}$ and $v_i$,) and non-physics entities ($d_{ij}^0$ and $v_i^0$) are mutual in nature: $d_{ij}^0$ and $v_i^0$ leads to change of $d_{ij}$ and $v_i$, which means opinions can change physical behavior, and the physics things can also change opinions inversely such that $d_{ij}$ and $v_i$ may also inversely change $d_{ij}^0$ and $v_i^0$ through perception and cognition.

In a sense the oscillation phenomenon is the origin of wave. Can we directly justify oscillation phenomenon of pedestrian crowd in reality? Sometimes it is observable for high-density crowd by using video-based technology, but usually it is a sign of crowd tragedy like stampede. "It was like a huge wave of sea gushing down on the pilgrims" (P. K. Abdul Ghafour, Arab News), this is the crowd disaster in Mena on January 12, 2006. At a passageway when the crowd become more and more impatient, the conscious wave exiting in $d_{ij}^0$ and $v_i^0$ will become physical, and people thus start to push others in order to move forward. Such pushing force generated a pressure wave propagating in the crowd, and at a certain point, usually considered as a bottleneck, the wave may stop and it shows the power of destruction: someone may fall down and stampede happens there. In brief, as mentioned in Helbing et al., 2002, the social-force model may symbolizes various phenomena of human society at the macroscopic scale, and we think that such oscillation has a more profound meaning to explore in several aspects.

## IV. Faster-is-Slower Effect and Yerkes–Dodson Law

Very interestingly, stress is actually a term which has dual meaning in both physics and psychology. It was initially a physical quantity, which is a measure of the internal forces in a body between its particles. In the 1920s and 1930s, biological and psychological circles occasionally used the term to refer to a mental strain or a harmful environmental agent that could cause illness. The current usage of the word stress arose out of Selye's 1930s experiments, where the term was referred not just to the harmful agent but to the state of the organism as it responds and adapts itself to the environment.

By searching in literature of physics, social psychology and emergency egress, we think that "stress" is more accurate conceptualizations of the social-force model (Sime, 1980; Ozel, 2001). Psychological stress can be understood as the interaction between the environment and the individual (Selye, 1978, Staal, 2004), emphasizing the role of the individual's appraisal of situations in shaping their responses. In our model such stress is the result of mismatch between psychological demand and realistic situation, and Equation (2) and (5) characterizes the mismatch in terms of velocity and distance: the psychological demand is represented by desired velocity $\boldsymbol{v}_i^0$ and desired interpersonal distance $d_{ij}^0$, while the physical reality is described by the physical velocity $\boldsymbol{v}$ and physical distance $d_{ij}$. The gap of two variables describes how much stress people are bearing in mind, and thus are motivated into certain behavior in reality. Such behavior is formulated as the self-driving force and social force in Equation (2) and (5).

Furthermore, velocity is a time-related concept in physics and the gap of velocities actually describes a kind of time-related stress, or commonly known as time-pressure. Such a kind of stress is caused by insufficient time when people are dealing with a time-critical situation, and time is the critical resource to complete the task. Besides, there is another kind of stress originating from social relationship and social space, and it may lead to competition or cooperation in crowd dynamics, and it refers to interpersonal distance and proxemics. Such stress is space-related, and is characterized by the social-force.

*Table 2. On Conception of Stress in Social-Force Model*

| | *Difference between subjective opinion and objective reality* |
|---|---|
| *Time-Related Stress: Velocities* | $\boldsymbol{v}_i^0 - \boldsymbol{v}_i$ |
| *Space-related Stress: Distances* | $d_{ij}^0 - d_{ij}$ |

The social-force model exhibits several spatio-temporal phenomena of crowd dynamics. Next, we will discuss a scenario in detail to justify conception of stress. This scenario was named by "faster-is-slower" effect in Helbing, Farkas and Vicsek, 2000, and it refers to egress performance when a large number of individuals pass through a narrow doorway. The simulation result shows that the egress time may inversely increase if the average desired velocity keeps increasing. In other words, egress performance may degenerates if the crowd desire moving too fast to escape. We will explain the simulation result from the psychological perspective of stress and time-pressure. In particular, this scenario reiterates an existing psychological knowledge: moderate stress improves human performance (i.e., speeding up crowd motion); while excessive stress impairs their performance (i.e., disorders and jamming), and this theorem is widely known as Yerkes–Dodson law in psychological study (Yerkes and Dodson, 1908; Teigen, 1994; Wikipedia, 2016).

Yerkes–Dodson law states the relationship between arousal level and performance: performance increases with arousal, but only up to a point. Beyond the point the arousal becomes excessive and the situation is much stressful such that performance diminishes. The arousal level indicates the intensity of motivation and it depends on stimulus strength from environment (e.g. alarm or hazard). Motivation leads to behavioral response. In the social-force model, the arousal or motivation is represented by desired velocity $\boldsymbol{v}^0$, and the behavioral response is represented by actual velocity $\boldsymbol{v}$. The performance of crowd escape is measured by pedestrian flow $\rho v$ at the doorway, describing how many individuals pass through a doorway of unit width per time unit (See Figure 6, $\rho$ and $v$ are the crowd density and physical speed nearby the doorway). The pedestrian flow is limited by the passage capacity, which determines the maximal pedestrian flow that people are able to realize in collective motion (Wang et al., 2008). In other words, the passageway capacity determines the critical point in Yerkes–Dodson law, indicating whether the collective motivation is excessive or not.

(a) When the passage capacity is sufficient, $v$ increases along with $v^o$ while $\rho$ can be adjusted such that the physical distance among people is psychologically comfortable. As a result, people are able to move as fast as desired while still keep proper interpersonal distance. This scenario corresponds to the increasing segment of the curve in Figure 6.

(b) When the passage is saturate, the physical speed $v$ and density $\rho$ reach the maximum and the pedestrian flow $\rho v$ is the maximal (IMO, 2007). In this situation further increasing $v^o$ will compress the crowd and increase the repulsion among people. As the repulsion increases, the risk of disorder and disaster at the bottleneck increases correspondingly (e.g., jamming and injury). If such disastrous events occur, the moving crowd will be significantly slowed down and the faster-is-slower effect comes into being, and this corresponds to the decreasing segment of the curve in Figure 6.

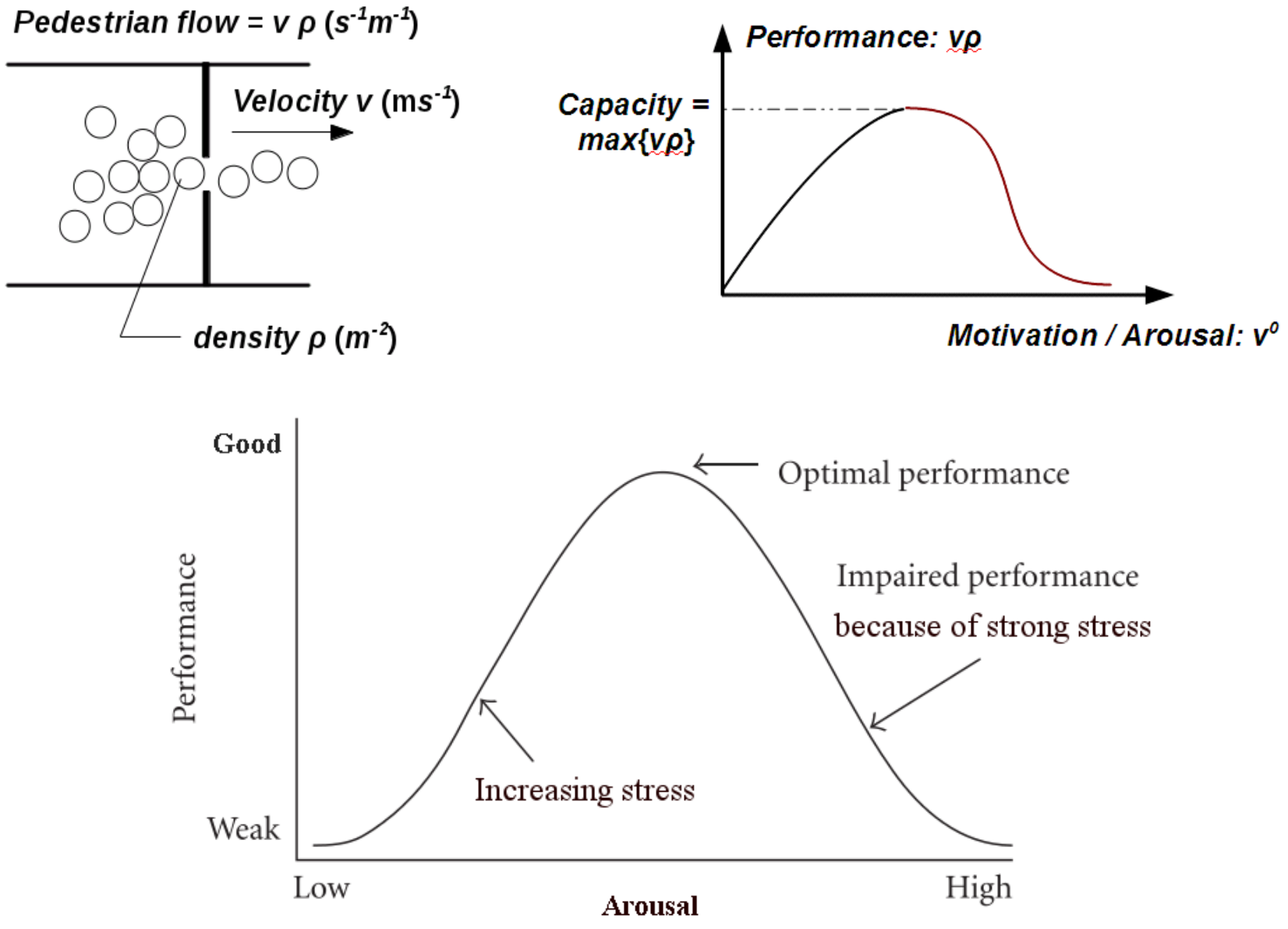

**Figure 6. About crowd movement at a bottleneck: The relationship between $v^0$ and performance $\rho v$ is depicted by an inverted-U curve, which is commonly known as Yerkes–Dodson law in psychological study. The ascending segment implies that faster-is-faster effect also exists as motivation level $v^0$ increases initially, and faster-is-slower effect emerges in the descending segment of the curve if $v^0$ keeps increasing and exceeds a certain limit.**

In sum, as motivation level $v^0$ increases, there are two scenarios as introduced above. The relationship between $v^0$ and performance $\rho v$ is depicted by an inverted-U curve as shown in Figure 6. Here the motivation level $v^0$ especially depends on environmental stressors, which are any event or stimulus perceived as threats or challenges to individuals. For example, in emergency evacuation a sort of important stressors are hazardous condition (e.g., fire and smoke). Perception of hazard will increase the arousal level so that the desired velocity $v^0$ increases.

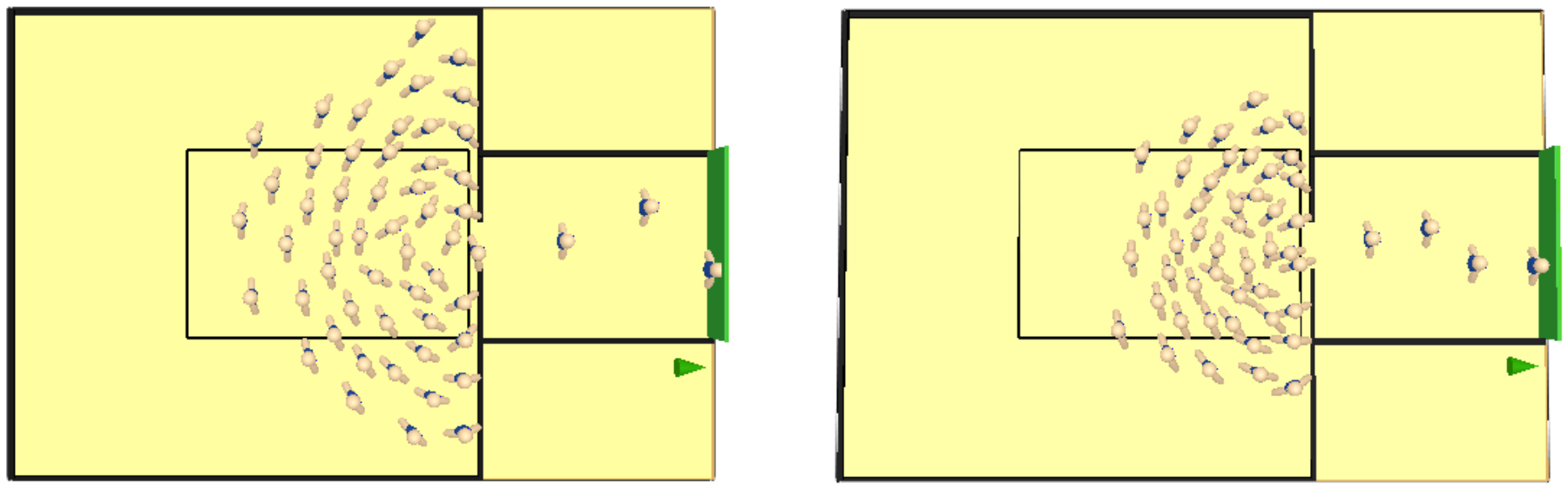

(a) (b)

**Figure 7. About Social Force and Faster-Is-Slower Effect: (a) Use large $d_{ij}^0$ in normal situation such that people obey social norm of large interpersonal distance. The result is decrease of flow rate and less chance of physical interaction. (b) Use small $d_{ij}^0$ in emergency egress such that people follow the social norm of small interpersonal distance. Flow rate thus increases and physical interaction also increase in a stochastic sense.**

The concept of desired interpersonal distance agrees with another psychological concept of social norms. Social norms are defined as "representations of appropriate behavior" in a certain situation or environment, and it defines proximal distance in various occasions. For example in elevators or entrance of a passageway, people commonly accept smaller proximal distance. Thus, the desired interpersonal distance is smaller and $d_{ij}^0$ is to be scaled down proportionally in these places. In emergency escape The interpersonal distance is also smaller than in normal situation, and people need to talk and exchange information with each other in emergency egress. The social norm is thus modified such that $d_{ij}^0$ is scaled down properly. The parameter of $A_i$ and $B_i$ may also be scaled down so that the social force as a whole is reduced in such an occasion (Korhonen, 2019).

A common result of scaling down $d_{ij}^0$ is that competitive behavior may emerge in crowd. In other words the physical force becomes effective among people and they may have more physical interaction at bottlenecks. As physical force is intensified, someone may fall down. The falling-down people become obstacle to others and thus slow down the egress flow, and they may cause others to fall down and this is so-called stampede disaster in crowd event. In sum the social force model with $d_{ij}^0$ is useful to investigate crowd behavior when falling-down effect is jointly applied. As below FDS+Evac is used to realize the falling-down effect where a pedestrian falls down if the physical force exceeds a threshold (Korhonen 2019; Forney, 2020).

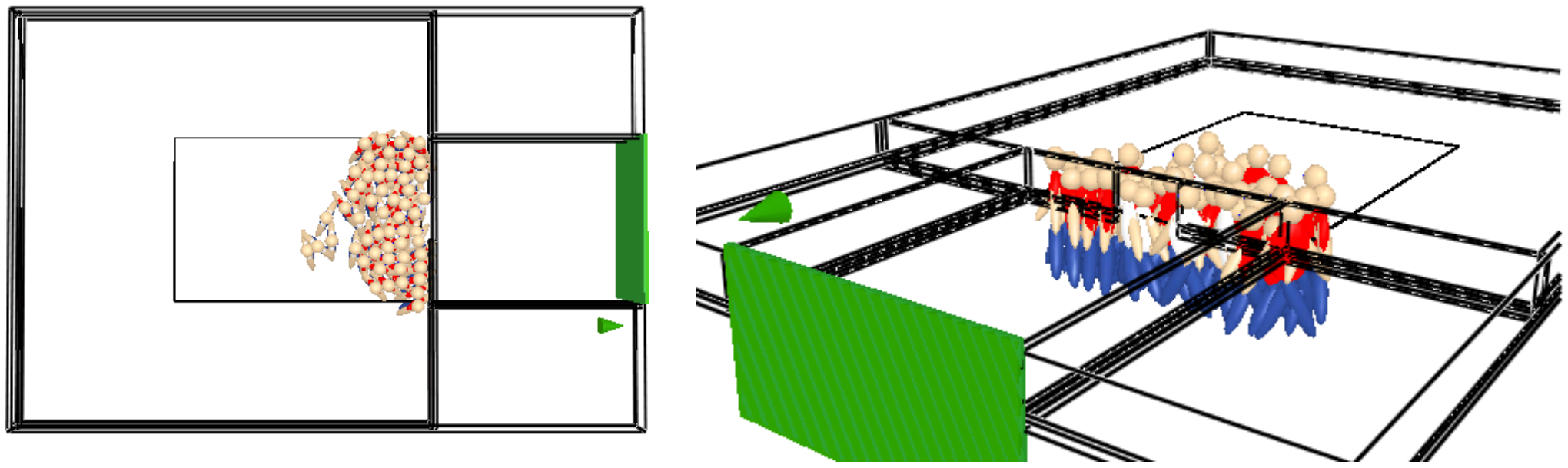

**Figure 8. Crowd Escape at Bottleneck with Falling-Down Model: The white agents are falling-down agent who cannot move and are considered as obstacle to the moving agents. They fall down because the physical force exceeds a given threshold. The red agents are moving agent toward exit, and they cannot simply get over the white ones to reach the door and the pedestrian flow rate is thus decreased.**

In sum, mismatch of psychological demand and physical reality results in a stressful condition. In emergency egress, such stress is aroused from environmental factors in two categories. A major kind of factors include hazard conditions and alarm, resulting in impatience of evacuees and causing time-pressure. The psychological model refers to the fight-or-flight response (Cannon, 1932), where hazardous stimulus motivate organisms to flee such that the desired velocity increases. Another kind of stress is aroused from surrounding people, resulting in interpersonal stress in collective behavior. Such stressor reduces interpersonal space so that people tend to stay together to generate a sense of safety. However, at bottlenecks competitive behavior in reduced space may cause disorder or disaster event. In brief, the simulation result about the faster-is-slower effect reiterate Yerkes–Dodson law with respect to two-dimensional stressors.

In psychology literature Yerkes–Dodson law is also understood by dividing stress into eustress and distress (Selye, 1975): stress that enhances function is considered eustress. Excessive stress that is not resolved through coping or adaptation, deemed distress, may lead to anxiety or withdrawal behavior and degenerate the performance. Thus, stress could either improve or impair human performance. Traditionally, this psychological theorem mainly refers to performance at individual level, such as class performance of a student or fight-or-flight response of an organism. The simulation of social-force model reiterates this well-known psychological knowledge in the sense of collective behavior. In brief, the testing result of social-force model agrees with Yerkes–Dodson law and it provides a new perspective to understand this classic psychological principle.

## CONCLUSIONS

The social-force model exhibits several spatio-temporal phenomena of crowd dynamics, but some controversial issues have been long existing on this model. To deal with the controversy we introduce a new concept of desired interpersonal distance to renew the formula of the social force, and it is a counterpart of the desired velocity in the self-driving force. Based on this new concept we explain that the model agrees with the Newton third law when it is applied to simulating pedestrian motion, and further provide a new perspective to understand oscillation phenomenon in existing simulation. The social-force model thus exhibits a bridge between the physics laws and psychological principles regarding crowd motion. A new angle is further presented to understand the model based on psychological concept of stress, where the self-driving force and social force reflect time-related stress and space-related stress, respectively. The faster-is-slower effect is thus reiterated by psychological discipline of Yerkes–Dodson law, which explains that excessive stress degenerate the performance of human behavior.

## Acknowledgments

The author is also grateful to Timo Korhonen and Mohcine Chraibi for discussion in simulation work. The author thank Dirk Helbing and Illes Farkas for helpful comments on this manuscript. The author is thankful to Peter Luh and Kerry Marsh for helpful comments on earlier work in University of Connecticut. The author appreciates the research program funded by NSF Grant # CMMI-1000495 (NSF Program Name: Building Emergency Evacuation - Innovative Modeling and Optimization).

## References

[1] M. Chraibi, U. Kemloh, A. Seyfried, A. Schadschneider, "Force-based models for pedestrian dynamics," Networks and Heterogeneous Media (NHM) , Vol. 6, No. 3, 2011, pp. 425-442.

[2] W. B. Cannon, *Wisdom of the Body*. United States: W.W. Norton & Company, 1932.

[3] G. P. Forney, "Smokeview, A Tool for Visualizing Fire Dynamics Simulation Data, Volume I: User's Guide", NIST Special Publication 1017-1 6th Edition, National Institute of Standards and Technology, Gaithersburg, MA, June 2020, 188 p.

[4] E. T. Hall. "A System for the Notation of Proxemic Behavior". American Anthropologist. Vol. 65, No. 5, pp. 1003–1026, 1963.

[5] E. T. Hall, The Hidden Dimension. Anchor Books. 1966.

[6] D. Helbing, Traffic and Related Self-Driven Many-Particle Systems, Reviews of Modern Physics, Vol. 73, No. 4, 2001, pp. 1067-1141.

[7] D. Helbing, L. Buzna, A. Johansson, and T. Werner, "Self-organized pedestrian crowd dynamics: Experiments, simulations, and design solutions." Transportation Science, 39(1): 1-24, 2005.

[8] D. Helbing, I. Farkas, and T. Vicsek, "Simulating Dynamical Features of Escape Panic," Nature, 407: 487– 490, 2000.

[9] D. Helbing, I. Farkas, P. Molnar, T. Vicsek, Simulation of pedestrian crowds in normal and evacuation situations, in: Schreckenberg, M., Sharma, S.D. (Eds.), Pedestrian and Evacuation Dynamics, Springer. pp. 21–58. 2002.

[10] D. Helbing and P. Molnar, "Social force model for pedestrian dynamics," Physical Review E, 51(5): 4282-4286, 1995.

[11] IMO, "Guidelines for Evacuation Analyses for New and Existing Passenger Ships", MSC/Circ.1238, International Maritime Organization, London, UK, 30 October 2007.

[12] A. Johansson, D. Helbing, and P. K. Shukla, Specification of the social force pedestrian model by evolutionary adjustment to video tracking data, Advances in Complex Systems, Vol. 10, pp. 271-288, 2007.

[13] A. Johansson, D. Helbing, H. Z. A-Abideen, and S. Al-Bosta, "From crowd dynamics to crowd safety: A video-based analysis," Advances in Complex Systems, 11(4): 497-527, 2008.

[14] T. Korhonen, Technical Reference and User's Guide for Fire Dynamics Simulator with Evacuation, (FDS+Evac, FDS 6.5.3, Evac 2.5.2), VTT Technical Research Center of Finland, 2019.

[15] T. Korhonen, S. Hostikka, S. Heliövaara, and H. Ehtamo, "FDS+Evac: An Agent Based Fire Evacuation Model", Proceedings of the 4th International Conference on Pedestrian and Evacuation Dynamics, February 27–29, 2008, Wuppertal, Germany.

[16] T. Korhonen and S. Hostikka, "Technical Reference and User's Guide for Fire Dynamics Simulator with Evacuation, FDS+Evac, (FDS 5.5.0, Evac 2.2.1)," VTT Technical Research Center of Finland, May. 2010.

[17] T. Kretz, "On Oscillations in the Social Force Model," Physica A: Statistical Mechanics and its Applications 438:272–285, 2015.

[18] K. Lewin, Field theory in social science, Harper, New York, 1951.

[19] Oasys Inc., The Verification and Validation of MassMotion for Evacuation Modelling, Issue 01, January 2018.

[20] F. Ozel, "Time Pressure and Stress as a Factor During Emergency Egress," Safety Science, 38: 95-107, 2001.

[21] H. J. Payne, Models of freeway traffic and control, in Mathematical Models of Public Systems, Vol. 1 of Simulation Councils Proc. Ser., pp. 51-60, 1971.

[22] N. Pelechano and N. I. Badler , "Modeling Crowd and Trained Leader Behavior during Building Evacuation," IEEE Computer Graphics and Applications, 26(6): 80-86, November-December 2006. Also posted at ScholarlyCommons@Penn. http://repository.upenn.edu/cis papers/272

[23] G. Proulx, "A Stress Model for People Facing a Fire," Journal of Environmental Psychology, 13(2): 137-147, 1993.

[24] J. D. Sime, "The Concept of Panic," Fires and Human Behavior, D. Canter (ed.), First edition, John Wiley & Sons, pp. 63-81, 1980.

[25] M. A. Staal, "Stress, Cognition, and Human Performance: A Literature Review and Conceptual Framework (NASA/TM – 204-212824)," August 2004, Hanover, MD: NASA Scientific and Technical Information Program Office.

[26] A. F. Stokes and K. Kite, "On grasping a nettle and becoming emotional." In P.A. Hancock, & P.A. Desmond (Eds.), Stress, workload, and fatigue. 2001, Mahwah, NJ: L. Erlbaum.

[27] K.H. Teigen, Yerkes-Dodson: A law for all seasons. Theory & Psychology, 4: 525-547, 1994.

[28] P. Wang, "Understanding Social Force Model in Psychological Principles of Collective Behavior" arXiv:1605.05146v9, 2017.

[29] P. Wang, P. B. Luh, S. C. Chang and J. Sun, "Modeling and Optimization of Crowd Guidance for Building Emergency Evacuation," Proceedings of the 2008 IEEE International Conference on Automation Science and Engineering (CASE 2008), Washington, D.C., pp. 328 – 334, August 2008.

[30] G.B. Whitham, Linear and nonlinear waves, John Wiley and Sons, New York, 1974.

[31] W.M. Winton, Do introductory textbooks present the Yerkes-Dodson law correctly? American Psychologist, 42: 202-203, 1987.

[32] R. M. Yerkes, J. D. Dodson. "The relation of strength of stimulus to rapidity of habit-formation." Journal of Comparative Neurology and Psychology, 18: 459–482, 1908.